\documentclass[sigconf]{acmart}

\usepackage{acronym}
\usepackage[inline]{enumitem}

\usepackage{todonotes}

\usepackage{cleveref} % This must be loaded LAST!

\acrodef{api}[API]{Application Program Interface}
\acrodef{ccps}[CCPE]{collective cyber-physical ecosystem}
\acrodef{CDN}{Content Delivery Network}
\acrodef{ci}[CI]{collective intelligence}
\acrodef{cci}[CCI]{computational collective intelligence}
\acrodef{CLR}{Common Language Runtime}
\acrodef{cps}[CPS]{cyber-physical system}
\acrodef{CS}{Collective System}
\acrodef{DSL}{domain-specific language}
\acrodef{DNS}{Domain Name System}
\acrodef{ECC}{edge-cloud continuum}
\acrodef{GPL}{General-Purpose Language}
\acrodef{hmi}[HMI]{Human-Machine Interface}
\acrodef{iot}[IoT]{Internet of Things}
\acrodef{JVM}{Java Virtual Machine}
\acrodef{se}[SE]{software engineering}
\acrodef{RL}{reinforcement learning}
\acrodef{rv}[RV]{Runtime Verification}
\acrodef{VM}{Virtual Machine}

\AtBeginDocument{%
  }

\setcopyright{none}
\copyrightyear{2024}
\acmYear{2024}
\acmConference[SE 2030]{International Workshop on Software Engineering in 2030}{November 2024}{Puerto Galinàs (Brazil)}
\sloppypar

\begin{document}

\title[Software Engineering for Collective Cyber-Physical Ecosystems]{Software Engineering for  Collective Cyber-Physical Ecosystems % on Computing Continua}
}

%\author{Roberto Casadei}
%\orcid{0000-0001-9149-949X}
%\email{roby.casadei@unibo.it}
%
%\author{Gianluca Aguzzi}
%\orcid{0000-0002-1553-4561}
%\email{gianluca.aguzzi@unibo.it}
%
%\author{Danilo Pianini}
%\orcid{0000-0002-8392-5409}
%\email{danilo.pianini@unibo.it}
%
%\author{Mirko Viroli}
%\orcid{0000-0003-2702-5702}
%\email{mirko.viroli@unibo.it}
%
%\affiliation{%
%  \institution{University of Bologna}
%  \streetaddress{50 Via dell'Università, 47521}
%  \city{Cesena}
%  \country{Italy}
%}
%
%\author{Giorgio Audrito}
%\orcid{0000-0002-2319-0375}
%\email{giorgio.audrito@unito.it}
%
%\author{Ferruccio Damiani}
%\orcid{0000-0001-8109-1706}
%\email{ferruccio.damiani@unito.it}
%
%\author{Giordano Scarso}
%\orcid{0009-0009-2114-7435}
%\email{giordano.scarso@unito.it}
%
%\author{Gianluca Torta}
%\orcid{0000-0002-4276-7213}
%\email{gianluca.torta@unito.it}
%
%\affiliation{%
%  \institution{University of Turin}
%  \streetaddress{185 Corso Svizzera, 10149}
%  \city{Turin}
%  \country{Italy}
%}

\author{Roberto Casadei}
\orcid{0000-0001-9149-949X}
\affiliation{%
  \institution{Università di Bologna}
  \streetaddress{50 Via dell'Università, 47521}
  \city{Cesena}
  \country{Italy}
}
\email{roby.casadei@unibo.it}

\author{Gianluca Aguzzi}
\orcid{0000-0002-1553-4561}
\affiliation{%
  \institution{Università di Bologna}
  \streetaddress{50 Via dell'Università, 47521}
  \city{Cesena}
  \country{Italy}
}
\email{gianluca.aguzzi@unibo.it}

\author{Giorgio Audrito}
\orcid{0000-0002-2319-0375}
\affiliation{%
  \institution{Università di Torino}
  \streetaddress{185 Corso Svizzera, 10149}
  \city{Turin}
  \country{Italy}
}
\email{giorgio.audrito@unito.it}

\author{Ferruccio Damiani}
\orcid{0000-0001-8109-1706}
\affiliation{%
  \institution{Università di Torino}
  \streetaddress{185 Corso Svizzera, 10149}
  \city{Turin}
  \country{Italy}
}
\email{ferruccio.damiani@unito.it}

\author{Danilo Pianini}
\orcid{0000-0002-8392-5409}
\affiliation{%
  \institution{Università di Bologna}
  \streetaddress{50 Via dell'Università, 47521}
  \city{Cesena}
  \country{Italy}
}
\email{danilo.pianini@unibo.it}

\author{Giordano Scarso}
\orcid{0009-0009-2114-7435}
\affiliation{%
  \institution{Università di Torino}
  \streetaddress{185 Corso Svizzera, 10149}
  \city{Turin}
  \country{Italy}
}
\email{giordano.scarso@unito.it}

\author{Gianluca Torta}
\orcid{0000-0002-4276-7213}
\affiliation{%
  \institution{Università di Torino}
  \streetaddress{185 Corso Svizzera, 10149}
  \city{Turin}
  \country{Italy}
}
\email{gianluca.torta@unito.it}

\author{Mirko Viroli}
\orcid{0000-0003-2702-5702}
\affiliation{%
  \institution{Università di Bologna}
  \streetaddress{50 Via dell'Università, 47521}
  \city{Cesena}
  \country{Italy}
}
\email{mirko.viroli@unibo.it}

\renewcommand{\shortauthors}{Casadei et al.}
\newcommand{\meta}[1]{{\color{blue}#1}}
\newcommand{\doneByFD}[1]{{\color{magenta}#1}}
\newcommand{\doneByGT}[1]{{\color{orange}#1}}

\begin{abstract}
Today's distributed and pervasive computing addresses large-scale cyber-physical ecosystems,
  characterised by dense and large networks of devices capable of computation, communication and interaction with the environment and people.
While most research focusses on treating these systems as ``composites'' (i.e., heterogeneous functional complexes), recent developments in fields such as self-organising systems and swarm robotics have opened up a complementary perspective: treating systems as ``collectives'' (i.e., uniform, collaborative, and self-organising groups of entities).
This article explores the motivations, state of the art,
  and implications of this ``collective computing paradigm'' in software engineering,
  discusses its peculiar challenges,
  and outlines a path for future research,
  touching on aspects such as macroprogramming,
  collective intelligence,
  self-adaptive middleware,
  learning, synthesis and
  experimentation of collective behaviour.
\end{abstract}



%  % \includegraphics[width=\textwidth]{sampleteaser}


\maketitle

\section{Introduction}

%\paragraph{General context.}
Technological advances in computing systems and networking
and trends like pervasive computing~\cite{DBLP:journals/percom/ContiDBKNPRTTZ12} and the \ac{iot}~\cite{DBLP:journals/isf/LiXZ15},
promote an increasing digitalisation of our world,
%
%Our planet, cities, and homes
% are
which is
being filled with interconnected computing devices
% creating a connection with the physical world and
 supporting a variety of services and applications.
In this work, we are concerned with the general high-level problem of
 how to unleash the potential of \emph{large-scale distributed systems}~\cite{DBLP:journals/jisa/SteenPV12}
by proper engineering of the software managing them or driving their overall behaviour.
%

%\paragraph{Specific context.} %Specifically,
Specifically, in this paper,
 we focus on \emph{\acp{ccps}}.  %\doneByFD{[NOTA LINGUISTICA: ho cambiato l'acronimo da CCPS a CCPE.]}
%
%We clarify what we mean by this term.
By \emph{cyber-physical},
 we indicate that systems consist of \emph{situated computing} devices
 able to sense and/or act in the physical environment;
 without neglecting that systems may also comprise non-situated and infrastructural devices
(e.g., edge servers, the cloud, etc.).
By \emph{ecosystem}, 
 we mean that devices \emph{live} and interact with each other within the environment,
 functioning as an ``ecological unit'', i.e., interacting to provide functionalities in a long-term equilibrium;
 in particular,
we consider both short- and long-lived tasks
which may require cooperation
and must adapt to perturbations.
By \emph{collective}, 
 we mean that multiple devices
 can be regarded as a whole,
 solving distributed tasks
 through \emph{\ac{cci}}~\cite{malone2022handbook-ci,DBLP:journals/alife/Casadei23,DBLP:journals/access/HePLMC19}.
Examples of \acp{ccps}
 include robotic swarms~\cite{DBLP:journals/swarm/BrambillaFBD13},
 smart cities~\cite{DBLP:journals/jzusc/DharmawanSFBW19},
 edge-cloud infrastructures~\cite{DBLP:journals/access/XuYGG18},
 and crowds of people with wearables~\cite{DBLP:conf/huc/FerschaLZ14}.

%\paragraph{A recent change in the research on SE for \acp{ccps}.}
%
In the research landscape of \emph{\ac{se}} for \acp{ccps},
 mainstream approaches -- like
 (micro-)services,
 programming frameworks,
 and artificial intelligence -- are typically applied
 under a view of \emph{``systems as composites''};
 in this view, a system is built as the \emph{integration} of \emph{heterogeneous} components,
 which can be engineered largely independently of the others.
Recently, results from niches of several research threads -- like
 collective adaptive systems~\cite{DBLP:journals/sttt/NicolaJW20},
 self-adaptive software~\cite{weyns2020self-adaptive},
 macro-programming~\cite{DBLP:journals/csur/Casadei23,DBLP:journals/jisa/JuniorSBP21},
 and swarm robotics~\cite{DBLP:journals/swarm/BrambillaFBD13} --  promoted
 the complementary view of \emph{``systems as collectives''}.
In this view,
 a system is engineered ``as a whole'',
 adopting (declarative) abstractions, tools, and methods
 suitable to coordinating groups of interacting devices,
 letting their \ac{cci} emerge.

%\paragraph{Contribution.}
%
In this paper,
 we address the impact of adopting a collective stance
in software engineering for large-scale distributed systems,
 considering \acp{ccps} as main target
 and artificial \ac{cci} as main goal.
To this end,
we provide a review of the state of the art on \ac{ccps} engineering,
 emphasising the different approaches involved
 and then highlight key challenges
 that we expect may contribute to the research and development of \ac{se}.

%\paragraph{Paper organisation.}
%
The rest of the manuscript is structured as follows.
\Cref{sec:motiv} details on the motivation for the emerging of the ``system as collectives'' perspective in software engineering.
%motivation.
%
\Cref{sec:sweng-collective} reports on the state of the art on \ac{ccps} engineering, and its main themes.
Then, in \Cref{sec:challenges},
 we highlight relevant challenges
 fostering research in software engineering for \acp{ccps}.
Finally, \Cref{sec:conc} offers a wrap-up.

\section{Systems as Collectives}\label{sec:motiv}

In this section, we detail on the motivation for a collective viewpoint in software engineering for large-scale distributed ecosystems.

\subsection{Context and high-level challenges}
%\paragraph{Context and high-level challenges.}
Engineering pervasive and cyber-physical (eco-)systems is challenging~\cite{DBLP:journals/computer/BroyS14,DBLP:journals/ccftpci/BeckerJLZ19}.
Indeed, many challenges have to be addressed,
 including
 the implications of \emph{distribution}
 (cf., lack of global clock, multiple administrative domains, dynamic topologies, latency and cost of communication, failure, inconsistency, security)~\cite{tanembaum-distrib-sys},
 the integration of \emph{discrete} and \emph{continuous} dynamics~\cite{cassandras2021intro-des},
 the implications of \emph{large scale} deployments~\cite{DBLP:journals/csur/OrgerieAL13},
 and the \emph{design of collaborative logic} (cf.~\Cref{sec:sweng-collective}).
For instance,
 addressing large scale typically requires
 the design of \emph{decentralised} solutions
 to avoid \emph{single points of failure} and \emph{performance bottlenecks}.
To deal with different requirements in terms of latency, cost, and performance,
 deployments often target multiple layers of the \emph{\ac{ECC}}~\cite{DBLP:journals/access/XuYGG18}.
Also,
 since large-scale deployments complicate \emph{maintenance},
 the design should involve \emph{autonomic capabilities}~\cite{DBLP:journals/computer/KephartC03},
 endowing systems with some \emph{adaptation} or \emph{self-* properties}
(e.g., self-organisation~\cite{DBLP:journals/alife/GershensonTWS20,DBLP:journals/tsmc/YeZV17},
self-reconfiguration~\cite{coullon2023swreconfig},
self-improvement and self-integration~\cite{DBLP:journals/fgcs/BellmanBDEGLLNP21}).

\subsection{SE approaches for distributed systems}
%\paragraph{Engineering approaches.}
%
The majority of current solutions for distributed software design
 leverage \emph{(micro-)services}~\cite{DBLP:books/sp/17/DragoniGLMMMS17},
 stream processing~\cite{DBLP:journals/jnca/AssuncaoVB18},
 or \emph{event-driven} platforms~\cite{DBLP:journals/tnsm/ChengZZC16}.
These techniques are suitable also for \acp{ccps}, though
 challenges related to heterogeneity, deployment, efficiency, integration, and security
 are hindering their adoption in  specific domains like manufacturing~\cite{DBLP:journals/access/XuYGG18}.
General approaches to \emph{service composition}~\cite{DBLP:journals/computer/Peltz03,DBLP:journals/csur/LemosDB16}
 include
 \emph{orchestration} by (possibly multiple) centralised entities~\cite{DBLP:journals/fgcs/VaqueroCEBSZ19}
 or \emph{choreographies} for decentralised interaction protocols~\cite{montesi2023introduction-choreographies}.
An interesting related \ac{se} practice
 is the use of \emph{architectural description languages}~\cite{DBLP:journals/sigsoft/Pandey10}
describing the components and constraints of an entire distributed system,
 and its specification in a single codebase---cf. \emph{multi-tier programming}~\cite{DBLP:journals/csur/WeisenburgerWS20}.

Concerning \emph{self-adaptive software systems}~\cite{weyns2020self-adaptive},
 there are multiple classes of self-adaptation approaches.
The \emph{reference model} for self-adaptive systems~\cite{DBLP:journals/computer/KephartC03} 
 distinguishes between the \emph{managing} and the \emph{managed} system,
 and generally organises a feedback loop around the \emph{MAPE-K (Monitor, Analyse, Plan, Execute - Knowledge)} components within the managing system.
Upon this basis, \emph{architecture-based} approaches
 emphasise runtime reasoning about architectural models;
 \emph{requirements-based} approaches
  emphasise the specification of requirements, often relaxed to deal with uncertainty, and meta-requirements about the self-adaptation itself;
  \emph{control-based} approaches leverage control theory to formally design and analyse the self-adaptation logic;
  and \emph{learning-based} approaches exploit machine learning to synthesise accurate models and plans.

In \emph{decentralised} self-adaptive systems,
 adaptation control is distributed among different components equipped with feedback loops.
This is similar but often distinguished from \emph{self-organising} systems~\cite{bonabeau1999swarm,DBLP:journals/tsmc/YeZV17},
 which generally consist of a larger number of simpler components that seek system goals by repeated local interaction.
Engineering self-organisation is a matter of identifying mechanisms and techniques to \emph{steer or guide the emergent behaviour} of groups of interacting agents~\cite{DBLP:journals/advcs/RodriguezGR07,DBLP:journals/tsmc/YeZV17}.

\subsection{From composites to collectives}\label{sec:a-change}
At this point, we deem useful to distinguish between
 two related but distinct kinds of possible target \emph{systems},
 where a system is generally meant, coherently with systems theories~\cite{systemsprinciples},
as a \emph{plurality} of entities considered together within an arbitrary boundary.
In particular, we are inspired by the ontological distinction between composites and collectives in Masolo et al.~\cite{DBLP:conf/fois/MasoloVFBP20}.
A \emph{composite}, a.k.a. an assembly or complex,
  consists of different kinds of components,
  thus exhibiting substantial \emph{heterogeneity},
  organised in a recursively decomposable %\footnote{
    %In other words, the component-hood relationship is transitive~\cite{DBLP:conf/fois/MasoloVFBP20}.
%}
  functional structure with multiple roles.
A distinct but related and non-disjoint notion
 is that of a \emph{collective},
 which is a more \emph{homogeneous} set of entities
 related by some non-transitive \emph{membership} relationship.
Roughly, then, examples of composites include the distributed software modules that collaboratively operate a manufacturing machine,
or a team of agents with significantly different capabilities interacting following a complex protocol;
 whereas a swarm of similar robots, a cluster, or a sensor network are exemplars of collectives.
Research in fields such as
swarm robotics~\cite{DBLP:journals/swarm/BrambillaFBD13},
collective adaptive systems~\cite{DBLP:journals/sttt/NicolaJW20},
 and
artificial life~\cite{DBLP:journals/alife/GershensonTWS20,DBLP:journals/swarm/BrambillaFBD13}
shows peculiar traits of collectives,
motivating new approaches to their engineering.
 Abowd~\cite{DBLP:journals/computer/Abowd16} refers to \emph{collective computing}
 as the fourth revolution after the mainframe, personal, and ubiquitous computing revolutions.
Indeed, the \emph{collective intelligence}~\cite{DBLP:journals/access/HePLMC19,DBLP:journals/csur/SuranPD20} emerging from the collaboration of humans and technologies has the potential to overcome important societal challenges~\cite{DBLP:journals/dgov/SuranPKHLKD22}.
Even when humans are not an active part of a solution,
 the engineering of \emph{artificial collectives}~\cite{DBLP:journals/alife/Casadei23}
 can contribute by improving existing applications (e.g., through increased autonomy and resilience),
 unlocking novel ones (cf. \emph{drone crowdsourcing}~\cite{DBLP:journals/access/AlwateerLF19} and \emph{\acs{iot}-as-a-Service}~\cite{DBLP:journals/iotj/HoqueHNIH22}),
 or sustaining the autonomous operation of large ecosystems in smart infrastructures and smart cities~\cite{DBLP:journals/fgcs/PianiniCVN21,DBLP:journals/jzusc/DharmawanSFBW19}.

The interest in large-scale \acp{ccps}
 is mounting.
In these systems,
 besides the local services and tasks (often resulting from \emph{complex networks of collaboration}),
 there is increasing interest
 in \emph{global-level services} (e.g., surveillance, waste collection, task allocation, group coordination)
 and \emph{global-level properties} (e.g., overall energy consumption, system-level resilience).
Accordingly, several research efforts
  are noticing \emph{fundamental gaps in software engineering for these systems}
  and consequently foster the
 \emph{collective viewpoint} in system analysis, design, and implementation,
 with \emph{collectives} and \emph{collective phenomena}~\cite{DBLP:journals/ao/WoodG09}
as first-class citizens~\cite{DBLP:journals/csur/Casadei23,DBLP:journals/sttt/MurgiaPTT23,DBLP:journals/jisa/JuniorSBP21,DBLP:conf/isola/InversoTT20}.
In the next sections, we detail on these contributions,
 further motivating interesting developments and challenges in software engineering.

\section{SE for CCPE$\mathbf{s}$: Overview}
\label{sec:sweng-collective}

In this section,
 we discuss the main issues encountered when engineering \acp{ccps}, %collective systems,
 and then
 provide a view of the state of the art.

\subsection{Main Themes in SE for CCPEs}
%\subsection{Main themes in engineering collective adaptive behaviour}

\subsubsection{Local-to-global and global-to-local mapping problems}
When dealing with \acp{ccps}, %collective adaptive systems,
  there are two main issues~\cite{wolpert1999introduction-ci-survey,SpatialIGI2013,DBLP:journals/alife/Casadei23}.

The \emph{local-to-global mapping problem}, also known as the \emph{forward} or \emph{prediction problem},
  entails determining what will be the global outcome
  out of an execution of the local behaviours making up the system, for a given set of environment dynamics.
Indeed, in the traditional ``node-centric'' engineering viewpoint,
 the programmer thinks about how the behaviour of an \emph{individual agent} will affect other agents and its surrounding environment.
However, to determine the global or system-level result,
 the behaviour of other agents and the environment have to be considered as well, resulting in a complex network of causes and effects, often leading to \emph{emergent phenomena}.
In emergence, it is difficult to trace the micro-level causes that led to a given macro-level observable.
In other words, emergence makes prediction difficult or impossible: the general solution consists of ``running the system'', i.e., seeing where the chain of events and the historical contingencies lead the system.
This makes \emph{simulation} an invaluable tool for studying and analysing collective behaviour.

The \emph{global-to-local mapping problem},
 also known as the \emph{inverse} or \emph{engineering problem},
 entails determining, from a %desired \\
 global outcome (\emph{target state}),
 what local behaviours will conduct to it and how.
Clearly the two problems are related, as two sides of the same coin.

\begin{figure}
\includegraphics[width=\columnwidth]{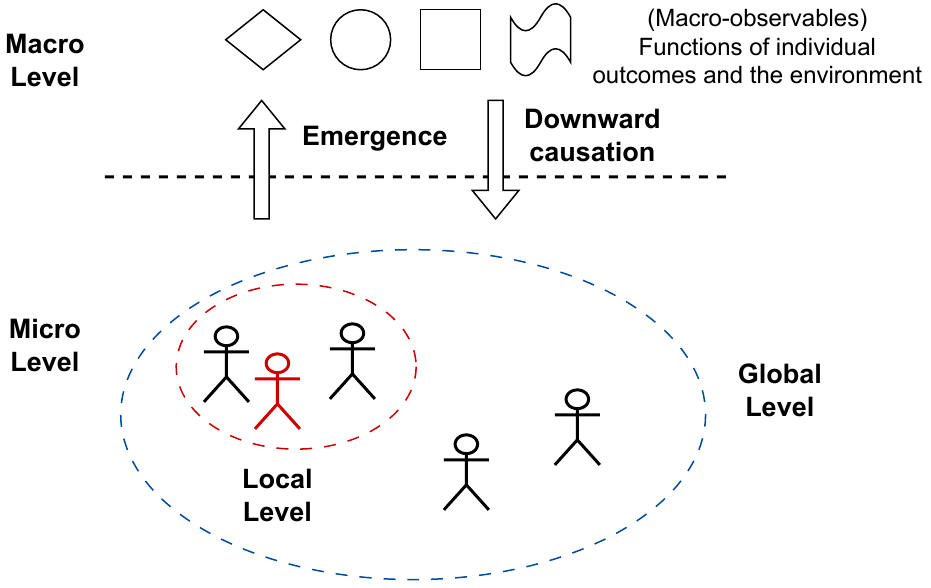}
\caption{The local/micro and global/macro levels of systems. The shapes at the top are just abstract denotations of macroscopic information obtained by an application-specific, abstracted function of the microscopic data taken from the agents and environment (cf.~\cite{DBLP:journals/access/HePLMC19,Gignoux2017}). The stylised agents are not necessarily people but also artificial autonomous agents.}
\label{fig:local-to-global-and-viceversa}
\end{figure}

Other names for these problems and corresponding solutions are also \emph{micro-to-macro} and \emph{macro-to-micro}, respectively~\cite{DBLP:journals/tcyb/YeCZLLW22,Sawyer2003micromacrolink,DBLP:journals/csur/Casadei23}.
It is relevant to introduce this terminology
 because several research threads do use these terms,
 and in our integrative effort we aim at removing fictitious linguistic barriers.
Indeed, in multi-agent systems, it is frequent to talk about the micro-level of agent activity and the macro-level of \emph{agent societies/organisations}~\cite{wooldridge2009intro-mas,Sawyer2003micromacrolink},
 and the research thread of \emph{macro-programming}~\cite{DBLP:journals/csur/Casadei23,DBLP:journals/jisa/JuniorSBP21}
 is exactly where the forward and inverse problems are addressed by using programming languages.
\Cref{fig:local-to-global-and-viceversa} depicts the general idea graphically,
 where macro-observables are abstracted as functions of the inputs from the individuals and environment,
 as in~\cite{DBLP:journals/access/HePLMC19,DBLP:series/lncs/NoelZ15}.

In the traditional ``node-centric'' engineering viewpoint,
  there is an implicit global-to-local solution
  that is limited to the designer's mind or at the requirements phase.
  In this paper, we analyze how recent research~\cite{DBLP:journals/sttt/NicolaJW20,DBLP:journals/csur/Casadei23,DBLP:journals/jisa/JuniorSBP21,SpatialIGI2013,DBLP:journals/software/PinciroliB16,DBLP:journals/csur/MottolaP11,DBLP:journals/swarm/BrambillaFBD13,DBLP:conf/isola/InversoTT20}
  shows that design can start
  from the ``global'' perspective,
  ideally with declarative definitions that represent the intended global behaviour,
  then implementing local behaviours by a formal global-to-local mapping technique,
  and finally going forward for proper testing and validation with a local-to-global mapping technique.

\subsubsection{Automatic vs.\ manual design}
Various works in the context of multi-agent systems and swarm robotics suggest that techniques for the design of collective behaviour can be classified in two main categories~\cite{DBLP:journals/swarm/BrambillaFBD13}: automated techniques and manual techniques.

\emph{Automated} design methods~\cite{DBLP:journals/amai/Buchberger23}
 work at the meta-level of algorithms,
 defining procedures for learning or synthesis
of the control programs driving (the individual agents of) \acp{ccps}. %collective systems.
Prominent examples of this family
 are \emph{evolutionary methods}~\cite{DBLP:series/sci/2008-108},
 \emph{multi-agent reinforcement learning}~\cite{DBLP:journals/tsmc/BusoniuBS08},
 and \emph{program synthesis}~\cite{DBLP:journals/ftpl/GulwaniPS17}.

\emph{Manual} or \emph{behaviour-based} design methods, instead,
 include the classical programming activity,
 where the programmer follows trial-and-error processes
 to craft algorithms using general-purpose or domain-specific languages.
When addressing collective behaviour,
 this is also referred to as \emph{macro-programming}~\cite{DBLP:journals/csur/Casadei23} (discussed more deeply in \Cref{sec:sota:macrop}).

Note that, being automated design
 the meta-level application of manual design,
 these two classes are not necessarily disjoint: indeed,
 key research opportunities lies in this overlap~\cite{DBLP:journals/ftpl/ChaudhuriEPSSY21,DBLP:conf/coordination/AguzziCV22,DBLP:conf/icdcs/AguzziCV22,DBLP:conf/nips/YangIBPSR21}.

\subsubsection{Understanding and building ``collective intelligence''}
\emph{\Acf{ci}} can be generally intended as the property of groups of individuals behaving in ways that seem intelligent~\cite{malone2022handbook-ci}.
This definition reuses possible definitions of ``intelligence'',
typically as a collection of task-specific skills
 or, in a stronger sense, as a general problem-solving and learning ability~\cite{chollet2019measure-intelligence}.
So, a \ac{ccps} %collective system
 is intelligent
 if it is able to solve tasks in a variety of environments,
 possibly adapting itself to cope with new problems.

%According to~\cite{DBLP:journals/access/HePLMC19},
% \ac{ci} can be understood according to three main paradigms of increasing complexity:
%(i) the \emph{isolation} paradigm, where the result is an aggregation of local behaviour results, with no interaction between the individuals and no feedback;
%(ii) the \emph{collaboration} paradigm, which features interaction between the individuals but the global outcome does not affect the individual behaviours;
%and
%(iii) the \emph{feedback} paradigm, which features both interaction and feedback, i.e., the global outcome affects the individual behaviours in turn.

%First of all, it is important to remark that
Remark that  the term ``collective intelligence''
  leads to two main related areas:
  socio-technical systems~\cite{DBLP:journals/csur/SuranPD20}
  and
  artificial collectives~\cite{DBLP:journals/alife/Casadei23}.
The former emphasises the complex collaboration of humans and technologies,
 whereas the latter emphasises the engineering of multi-agent systems consisting primarily of artificial devices.
%
%Of course, t
The two research areas are related, and humans are indeed considered in the latter kind of systems as well, but the problems addressed are quite different
 (roughly, more human-computer interaction on one hand,
  and more behaviour control design on the other hand).
Since socio-technical systems tend to feature increased complexity of agents and heterogeneity, hence leading to composites, in this work we mainly focus on artificial collectives.
%
% Since \ac{ci} often refers to natural systems
%  or socio-technical systems~\cite{malone2022handbook-ci,DBLP:journals/access/HePLMC19},
%  where computing technology supports intelligent human activity,
We explicitly refer to the interpretation of \ac{ci} as  \emph{\acf{cci}}~\cite{szuba2001cci,DBLP:journals/alife/Casadei23},
 meaning the intelligence globally exhibited by groups of computing devices.

\subsection{State of the Art in SE for CCPEs}
%\subsection{State of the art in Collective Systems Engineering}

It follows a non-exhaustive review of contributions to \ac{se} for \acp{ccps},
 aiming at illustrating its main problems, methods, and techniques.

\subsubsection{Learning/evolving collective behaviour}
The automatic approach in \acp{ccps} %collective system
  design has been a focal area for over 20 years,
encompassing various fields such as swarm robotics~\cite{DBLP:journals/swarm/BrambillaFBD13}, traffic management, and crowd engineering.
In this area, two main approaches have been studied: \emph{evolutionary approaches}~\cite{DBLP:series/sci/2008-108} and \emph{\ac{RL}}~\cite{Sutton2018-hc}.
Both rely heavily on simulation to provide necessary feedback for refining generated controllers.
Evolutionary approaches draw inspiration from natural evolution,
starting with a set of controllers that evolve based on simulation feedback.
The design of a controller is highly dependent on the scenario and desired complexity, using finite state machines~\cite{DBLP:journals/swarm/FrancescaBBTB14}, neural networks~\cite{Floreano2010}, and Boolean networks~\cite{DBLP:conf/wivace/BracciniRVS16}.
The state of the art in this research area is attributed to AuToMoDe~\cite{DBLP:journals/swarm/FrancescaBBGMPR15}, a general framework for generating controllers for robot swarms.

In the realm of \emph{collective \ac{RL}}~\cite{DBLP:journals/iotj/LiPYSLSZ22},
 initial efforts, such as those by COIN~\cite{coin},
 focused on bottom-up \ac{cci},
 aiming to enable multiple agents to learn concurrently via \ac{RL}.
These early efforts evolved towards prioritizing the collective as a primary entity, thus shifting focus to designing individual agent controllers.
For instance, a novel model named swarMDP~\cite{DBLP:journals/corr/SosicKZK16} was introduced to simplify
 the modelling of learning processes for swarm-like systems,
 characterised by homogeneous entities (same controller) with a partial observability of the environment.
This model represents an initial top-down approach,
 where the controller, possessing a global view,
 is then distributed into similar controllers for individual robots.

In the state of the art, there are two main strategies.
One strategy,
 aimed at collaborative systems with few agents
 and exemplified by COMA~\cite{DBLP:conf/aaai/FoersterFANW18} and MAPPO~\cite{DBLP:conf/nips/YuVVGWBW22},
 addresses the individual behaviour of agents that cooperate to achieve collective intelligence.
Another strategy,
 targetting more large-scale systems,
%and modern approaches where scale is essential,
 %and collectivity 
 let collective behaviour emerge through a uniform controller,
  akin to swarMDP but leveraging more recent models like mean-field games~\cite{DBLP:conf/icml/YangLLZZW18}---this is often known as \emph{many-agent} reinforcement learning~\cite{yang2021many}.
In the \emph{centralised training and decentralised execution} approach~\cite{DBLP:journals/air/GronauerD22,DBLP:journals/ijon/KraemerB16},
 controllers are derived from a global perspective and then distributed throughout the system.
 This methodology simplifies the learning of collective behaviours and allows for the verification of outcomes during simulation.

\subsubsection{Macroprogramming}\label{sec:sota:macrop}
Macroprogramming is the umbrella term that
 gathers programming solutions
 for defining and executing the macroscopic behaviour
 of software systems~\cite{DBLP:journals/csur/Casadei23,DBLP:journals/jisa/JuniorSBP21}.
The definition is voluntarily broad,
 since it mostly serves as a basis for
 integrating and finding commonalities among
 different solutions for similar problems
 recurring in disparate domains, from wireless sensor networks~\cite{DBLP:conf/ipsn/NewtonAW05,DBLP:journals/csur/MottolaP11} and parallel computing~\cite{giraphpp}
 to swarm robotics~\cite{DBLP:journals/software/PinciroliB16} and software-defined networks~\cite{DBLP:conf/conext/KangLRW13}.
So, the real question is not whether a programming solution is an instance of macroprogramming
 but rather to \emph{what extent it is so}~\cite{DBLP:journals/csur/Casadei23}.
In particular,  certain macro-programming approaches
 do expose \emph{collective-level abstractions},
  and programs can be read as instructions
  targetting not an individual device,
  but rather a collective of devices,
  hence \emph{the computing machine is conceptually the whole distributed system of computers}.
Achieving that, often entails some forms of \emph{declarativity} at the language level,
 a compilation of local programs from the global specification~(cf. \cite{DBLP:journals/csur/WeisenburgerWS20}),
 and a non-trivial interplay between the local/global programs
 and the middleware (cf. \Cref{subsec:abstractions}).

For a detailed picture of the state of the art on macroprogramming,
 the reader can refer to these two recent surveys~\cite{DBLP:journals/csur/Casadei23,DBLP:journals/jisa/JuniorSBP21}.
It should be noticed that the surveys do not cover
 research contributions that are rather focussed on smaller-scale heterogeneous systems, i.e., composites,
 and hence captured by the related fields of \emph{choreographic programming}~\cite{felipe2020choreographic-programming},
 \emph{multi-agent oriented programming}~\cite{boissier2020multi-agent-oriented-prog-mitpress},
 \emph{architectural description languages}~\cite{DBLP:journals/tse/MalavoltaLMPT13}, and
 \emph{multi-tier programming}~\cite{DBLP:journals/csur/WeisenburgerWS20}.

\subsubsection{Aggregate Computing}
Aggregate computing is a prominent macroprogramming approach for %collective adaptive systems
\acp{ccps}~\cite{DBLP:journals/jlap/ViroliBDACP19}, developed for more than ten years.
We briefly recall its fundamental features:
 \emph{bio-inspiration, formality, pragmatism, compositionality,  deployment-independence}.

Generally speaking, a major general theme in software engineering for \acp{ccps} is finding ways to map scientific theories about natural phenomena to software engineering constructs (e.g., programming abstractions or platforms)---cf. \emph{bio-inspired} mechanisms~\cite{DBLP:journals/nc/Fernandez-MarquezSMVA13}.
In the case of aggregate computing,
 the mechanism of self-organisation~\cite{DBLP:journals/alife/GershensonTWS20}
 is reflected in the execution model (implemented in the runtime or middleware) of programs written in this paradigm,
 which assumes that each device computes in repeated asynchronous sense--compute--interact rounds.

The programming paradigm is formally founded on \emph{field calculi}~\cite{Audrito-et-al:TOCL-2019},
 functional core languages
 for expressing manipulations of \emph{computational fields}: distributed data structures mapping sets of devices to values, hence denoting collective inputs and outputs.
This enables deriving guarantees and proofs about programs,
 e.g. to assess that a collective computation is \emph{self-stabilising} (roughly, eventually converging to the ``right'' value)~\cite{DBLP:journals/tomacs/ViroliABDP18}.

These calculi are implemented by standalone or embedded \emph{\acp{DSL}}~\cite{DBLP:journals/jlap/ViroliBDACP19},
 also providing pragmatic libraries of reusable functions capturing recurrent patterns of collective behaviour.
Using these languages and libraries, the programmer defines a single program expressing the whole behaviour of a collective as a \emph{composition} of building blocks.

Then, the program can be deployed on a network of devices, using a simulator or leveraging a proper middleware support on real systems~\cite{DBLP:journals/iotj/CasadeiFPPSV22}.
By distinguishing the logical model of the system
 from its physical deployment,
 it is also possible to partition an application
 into deployable components
 and hence offload computations across the edge-cloud continuum~\cite{DBLP:journals/iotj/CasadeiFPPSV22}.

\subsubsection{Models, Methods, and Tools for \acp{ccps}}

A few works provide early contributions
 addressing (parts of) the development life-cycle
 for \acp{ccps}, mainly from the research area of collective adaptive systems~\cite{DBLP:journals/sttt/NicolaJW20}.
Actually, the majority of these, briefly reviewed in the following, lie at the intersection of composites and collectives.

There is a number of contributions about the use of formal methods for the specification and quantitative evaluation of collective adaptive systems~\cite{DBLP:conf/sfm/2016}.
One example is \emph{CARMA (Collective Adaptive Resource-sharing Markovian Agents)}~\cite{DBLP:conf/sfm/LoretiH16}, a stochastic process algebra
  supporting the modelling and verification,
  through a stochastic simulator,
  of quantitative properties of collective behaviour like performance, availability, dependability.

Another major source of contributions was the \emph{ASCENS (Autonomic Service-Component ENSembles)} project~\cite{ascens-project}.
This project proposed approaches and formal languages for the specification and analysis of \emph{ensembles}, namely dynamic groups of devices exhibiting complex interactions and working in complex environments~\cite{DBLP:conf/fmco/WirsingHTZ11}.
In the \emph{SOTA (State Of The Affairs)} and \emph{GEM (General Ensemble Model)} approach~\cite{DBLP:journals/sttt/AbeywickramaBMZ20},
  collective behaviour is denoted as a trajectory in a state space where each point denotes a single ``state of the affairs'' comprising all the information about the ensemble and its environment.
Regarding the programming of ensembles, ASCENS features the \emph{SCEL (Service Component Ensemble Language)}~\cite{DBLP:journals/taas/NicolaLPT14},
 a process-algebraic abstract language
 for specifying behaviour of individual entities,
 of aggregations of entities,
 based on \emph{attribute-based communication}~\cite{DBLP:journals/scp/AlrahmanNL20}, and
 in a way that is parametric to knowledge repositories and adaptation policies.

A related approach to engineering dynamic ensembles is provided by Bucchiarone et al.~\cite{DBLP:conf/birthday/BucchiaroneM19},
 based on \emph{domain objects} encapsulating behavioural processes
 and exporting \emph{process fragments} for refinement and composition.
These networks of domain objects (i.e., ensembles)
 can be modelled using typed graph grammars~\cite{DBLP:conf/sigsoft/HirschIM98}.

A programming framework for hybrid collaborative systems
 is proposed by Scekic et al.~\cite{DBLP:journals/tetc/ScekicSVRTMD20}.
There, collectives are teams of \emph{peers} (abstracting \emph{humans} and \emph{machines})
 performing ``collective-based tasks'', e.g. in crowdsourcing settings.
The approach, based on the SmartSociety framework~\cite{DBLP:journals/tetc/ScekicSVRTMD20},
 provides a Java-based platform and \ac{api} for the \emph{orchestration}
 of such hybrid ensembles,
 supporting \emph{managed} collectives,
 collective task management,
 adaptation policies (expressed in terms of changing handlers attached to task transitions),
 as well as other services for communication and incentive management.
The interesting part of the approach is the combination of \emph{collectiveness} (where collectives are the first-class citizens managed by the platform)
 and \emph{human orchestration},
 and the positioning of the contribution roughly half-way between composites and collectives engineering;
 however the approach is quite ad-hoc and with limited methodological guidance.

There are other approaches similar to SCEL, domain objects, and SmartSociety that share some of their motivation and abstractions, and provide related \acp{DSL} and platforms supporting modelling, implementation, and limited forms of verification. These include, e.g., \emph{DEECo (Distributed Emergent Ensembles of Components)}~\cite{DBLP:conf/cbse/BuresGHKKP13}.

Generally, any approach like aggregate computing, CARMA, or ASCENS comes with its own toolchain,
 but there exist also ``standalone'' tools.
One toolkit for reasoning about collective adaptive systems is \emph{Sibilla}~\cite{DBLP:conf/coordination/GiudiceMQRL22}.
It is a modular framework
 supporting multiple specification languages (e.g., for interactive objects, population models, and agent-based systems),
 organised with a three-layer architecture comprising:
 (i) a back-end supporting modelling, simulation, and analysis;
 (ii) different front-ends (\acp{api} and shells),
 and
 (iii) a runtime system coordinating activities.

\subsubsection{Methodologies for \acp{ccps}}

The aforementioned SOTA and GEM models
 promote \emph{goal-oriented}~\cite{DBLP:journals/re/HorkoffACLMPSPM19} adaptation requirements engineering.
However, there are also studies on specific properties of \acp{ccps} or peculiar aspects of their engineering.

In~\cite{DBLP:journals/fgcs/AliBKB22},
 a systematic review is carried out
 studying how existing approaches deal with \emph{uncertainty} and \emph{adaptation} in \acp{ccps}. %collective \acp{cps}.
Also, a \emph{design guide} is proposed for uncertainty-aware components in \acp{ccps},
 highlighting the need of identifying uncertainty sources  (e.g., failure, delays, noise, missing information) and ``border situations''
 to feed methods for handling uncertainties (e.g., humans-in-the-loop, reconfiguration, and declarative behaviour)
 before performing the adaptation decision-making.

Regarding engineering self-organisation and emergence, there exist some ideas and high-level guidelines~\cite{DBLP:series/lncs/NoelZ15,prokopenko2013guidedselforg}.
For instance, N{\"o}el and Zambonelli~\cite{DBLP:series/lncs/NoelZ15}
  contrast with approaches based on micro-level design
  and rather suggest to focus on the macro-level: \emph{``the global behaviour must emerge, exploiting self-organisation imposes
some design constraints and decomposition plays an important role in what can
emerge''}.
In particular, they delineate two different decomposition strategies:
  \emph{organisation-based} (by distinguishing roles)
  and
  \emph{functionality-based} (by breaking global tasks into global sub-tasks).
In the former strategy,
 literature on multi-agent organisational paradigms is especially relevant~\cite{DBLP:journals/ker/HorlingL04}.

\section{SE Challenges for CCPE$\mathbf{s}$}\label{sec:challenges}

In Section~\ref{sec:sweng-collective} we provided an overview
 of works contributing to the engineering
 of \acp{ccps}.
At their core,
 there are notations and tools supporting automatic and/or manual approaches
 helping to bridge the intended global behaviour
 with the local control program of the agents of the \ac{ccps} %collective system
   at hand.
The general goal is promoting forms of collective intelligence out of whole networks of devices. % making up the system.
The key insight is that several research works~\cite{DBLP:journals/sttt/NicolaJW20,DBLP:journals/csur/Casadei23,DBLP:journals/jisa/JuniorSBP21,SpatialIGI2013,DBLP:journals/software/PinciroliB16,DBLP:journals/csur/MottolaP11,DBLP:journals/swarm/BrambillaFBD13,DBLP:conf/isola/InversoTT20} %(across disparate domains)
 witness the need of embracing the collective viewpoint, possibly throughout the software engineering process~\cite{ascens-project,DBLP:journals/sttt/NicolaJW20}: cf.
 requirements expressing system-wide properties~\cite{DBLP:journals/sttt/AbeywickramaBMZ20},
 models and languages considering groups of entities and values as first-class abstractions~\cite{DBLP:journals/csur/Casadei23,DBLP:journals/jisa/JuniorSBP21,DBLP:conf/isola/InversoTT20},
 verification techniques leveraging model-checking or simulation to assess the emergents~\cite{DBLP:conf/sfm/LoretiH16,DBLP:conf/coordination/GiudiceMQRL22,DBLP:journals/tomacs/ViroliABDP18}.
In the following,
 we highlight gaps in the state of the art,
 denoting key challenges
 to be addressed in the context of software engineering
 of next-generation distributed systems.

%\subsection{Resolving the tension between homogeneity and heterogeneity}\label{sec:homo-hete}
\subsection{Handling Homogeneity and Heterogeneity}\label{sec:homo-hete}
%Domain-specific resolving of the tension between homogeneity and heterogeneity

\begin{figure}
    \includegraphics[width=\columnwidth]{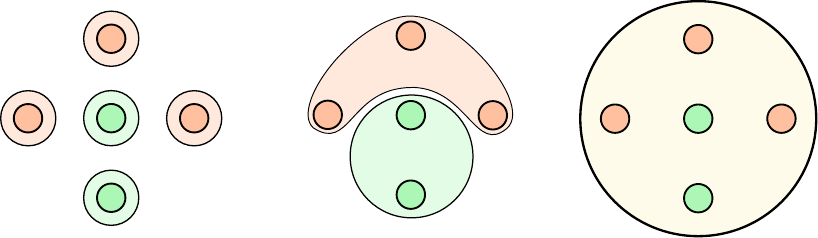}
    \caption{
        From composites to collectives.
        In composites (left) every entity composing the system must be observed, understood, and designed/trained individually;
        communication between entities needs to be modelled explicitly.
        In collectives (right) all entities are observed, understood, and designed/trained as a single entity;
        communication is implicit;
        dealing with the heterogeneity of the single entities is still an active research topic.
        Hybrid systems (centre) are possible:
        the system gets partitioned into multiple subsystems,
        typically capturing homogeneous devices, building a de-facto composite of collectives.
    }
    \label{fig:composites-collectives}
\end{figure}

\subsubsection{Integrating collectives and composites}
A major challenge in engineering \acp{ccps} is the integration of heterogeneous entities
into a coherent framework,
where the system can be inspected, understood, and designed (or trained)
as a single entity
(in the spirit of macroprogramming, cf. \Cref{sec:sota:macrop}).
Macroprogramming approaches, in fact,
are particularly well-suited for systems composed of homogeneous entities,
where the collective behaviour can be abstracted and reasoned about at a high level of abstraction.
Thus far, however,
there is a lack of formal and practical methods to deal with heterogeneity
from a collective perspective,
exploiting the peculiarities of every device without renouncing the benefit of
observing and designing
the system as whole.

In \Cref{fig:composites-collectives},
we provide a graphical representation of the difference between composites and collectives,
including also somewhat \emph{hybrid} systems,
where the whole composed of heterogeneous devices is partitioned into multiple subsystems
based on their internal homogeneity.
Although this idea helps with managing complexity,
it does not fully exploit the potential of a collective viewpoint
over a heterogeneous system.
Consider, for instance,
a system composed of drones and ground robots:
a truly collective approach would be able to form mixed teams
with capabilities that can only be achieved by combining the two types of devices
(for instance, exploiting the aerial view in specific areas to assist ground robots in navigation and path-finding);
while
a hybrid approach would partition the system into two homogeneous communicating subsystems,
losing the ability to create multiple small independent teams.

Providing means to tame heterogeneity without losing the privileged observation position of the collective
has great potential to enable a leap forward in several fields, including
swarm robotics~\cite{Dorigo2020, DBLP:conf/prima/KengyelHZRWS15},
the \ac{iot}~\cite{DBLP:journals/comsur/QiuCLAZ18, DBLP:conf/saso/CasadeiV18},
and the edge-cloud continuum~\cite{DBLP:conf/ccgrid/FerrerB0TK21}.

\subsubsection{Multi-targeting and software product lines}
The theme of heterogeneity,
largely impacting on how abstractions can be captured
(cf. \Cref{subsec:abstractions}),
has rippling effects on the software development process.
In fact, even though the software is designed from a collective perspective,
it will have to be executed on possibly very diverse devices,
ranging from edge servers and cloud instances to smartphones and wearables to
modest microcontrollers and sensors.
Although decompositional approaches have been proposed to tackle the problem of some devices
being unable to participate in some parts of the collective computation~\cite{DBLP:journals/fi/CasadeiPPVW20},
they do not account for the production of multiple versions of the software
that can execute on very diverse devices.
This problem can be fractioned into two macro sub-problems:
\begin{enumerate*}
    \item producing software \emph{compatible} with different target platforms from a single specification, and
    \item customising the specific software configuration considering the final target.
\end{enumerate*}
The first problem imposes a significant maintenance burden on potential middlewares
(as they must be able to run on multiple platforms),
potential external \acp{DSL}
(as they must be able to generate code for multiple targets),
and may severely limit the adoption of internal \acp{DSL}
(as they are constrained by the platforms targeted by the host language).
The latter problem calls for future \ac{ccps}-specific
software product lines~\cite{DBLP:journals/jss/BeekS23},
supporting the customisation of the final software in a target-specific fashion.

\subsection{Specifying and Running Collectives: %Abstractions,
\acp{DSL}, Middlewares, and Their Interplay}\label{subsec:abstractions}

Capturing collectives in a way that is actionable analysts and designers
requires appropriate abstractions to be reified.
Thus,
at some abstraction level, multiple devices must operate as a single one,
although in the respect of their possible heterogeneity.
Doing so, technically, can be performed in two ways:
\begin{itemize}
    \item through a middleware, creating a software layer that abstracts the collective as a single entity; or
    \item through a \ac{DSL},
    capturing the collective behaviour at design time and translating it into executable specifications.
\end{itemize}

Middlewares are the historically preferred approach
to building a shared layer multiple autonomous entities can interact through~\cite{DBLP:journals/tosem/MameiZ09,DBLP:journals/aai/PoggiRT02}.
Indeed, a middleware is a natural solution,
as it provides a shared unified layer, enforcing homogeneity over possibly heterogeneous devices.
This strategy has been used in the past for non-distributed system with success:
we may ascribe to the ``middleware'' category all those execution runtimes that could be targeted by multiple programming languages,
such as the \ac{JVM} (targeted by Java, Ruby, Scala, Kotlin, Groovy, and many other languages) or the
\ac{CLR} (targeted by VisualBasic.NET, C\#, F\#, \ldots).
On the other hand,
however,
middlewares can be impractical,
as they normally require multiple implementations for different platforms,
and they impose certain requirements on the underlying systems.
In the exemplary case of the \ac{JVM}
(which is akin in other runtimes)
the middleware has to be implemented for each supported platform,
with each implementation possibly featuring slightly different implementations.
We observe that there is a trade-off between complexity and portability:
the middleware's portability will be favoured by minimality;
on the other hand,
a minimal middleware will hardly provide high-level abstractions suitable for high-complexity systems,
thus requiring further layers,
each one possibly imposing further restrictions on the underlying platform.
For instance,
a middleware developed on top of the \ac{JVM} will not be capable to be executed on devices or runtimes for which a \ac{JVM} is not available,
such as the browser, or low-power wearables.

\acp{DSL} come instead with a different set of trade-offs.
The idea behind a \acp{DSL} is to create new abstractions as part of the language itself,
letting the compiler (interpreter) figure out how to translate (execute) the program.
\acp{DSL} can be realised in two ways:
as \emph{standalone} (or \emph{external}) languages~\cite{Riti2017external},
or \emph{embedded} (\emph{internal}) into an existing \ac{GPL}~\cite{Riti2017internal}.
The former approach is more flexible,
but it requires more development and maintenance effort,
while the latter is constrained by the syntax and semantics of the host \ac{GPL},
but, in turn, can immediately leverage its compiler/interpreter, its tooling,
and feature a much gentler learning curve for designers acquainted with the host \ac{GPL};
an internal \ac{DSL}, de facto, is a library designed to exploit the syntactic features of the host language
in such a way that the feeling is akin to using a dedicated language.

The two approaches are not mutually exclusive,
rather,
they reinforce each other.
In fact,
at some point,
the code written in a \ac{DSL} will perform communications over a network;
and these communications could well be mediated by a middleware.
On the other hand,
the infrastructural support a middleware provides will be exposed to the designers through an \ac{api},
which may be expressed, e.g., in form of an internal \ac{DSL}.
In practice, we have a spectrum of possibilities,
and understanding the better trade-off from the two extremes of a pure \ac{DSL}
directly calling networking primitives under the hood
and a pure middleware driven through a regular \ac{api}
is yet to be investigated.
This investigation becomes even more challenging under the consideration that
different applications
(with different contexts and objectives)
developed by different teams
(with different expertises)
may benefit from different balances,
and one specific design will hardly fit well in all conditions.
Clear guidance
directed to \ac{ccps} \ac{DSL} and middleware providers
on the trade-off each design choice brings on the table
is yet to be devised.

\subsection{Verification and Validation}

\subsubsection*{Static analysis.}
A first line of defence against bugs, smells, and vulnerabilities is static analysis;
however,
the adoption of static methods based on model checking is limited by the size, openness, and autonomy of \acp{ccps}~\cite{TestingCAS15}.
In response,
formal methods are being developed to support \ac{ccps}-tailored analysis~\cite{DBLP:conf/vmcai/StefanoL23,DBLP:journals/lmcs/CastiglioniLT23}.
Further steps are required in this direction,
and, from a practitioner's point of view,
these research efforts need to be reified into actionable tools
(possibly language- or middleware- specific).

\subsubsection*{Simulation.}
In distributed systems, testing is a main challenge~\cite{DBLP:journals/tcst/Rafajlowicz08}:
executing the software on a single machine
(even when software pieces are isolated through \acp{VM} or containers)
attenuates (or removes altogether) many of the potential issues to be faced at runtime,
such as network failures, slowdowns, \ac{CDN} updates, and internal \ac{DNS} lookup issues.
In the case of \acp{ccps},
the problem is exacerbated by the large scale,
which makes it unpractical (when plainly impossible) to test the software on a single machine.
The most immediate consequence is that \emph{simulation}
becomes a necessity in multiple phases of the software production~\cite{SimTestRobots21}:
fast simulator with a very simplified model of the world can support the developer in the early stages of development;
while more detailed simulations, with accurate physics and network models (and much longer time to execute)
are needed to support in-depth testing~\cite{TestingADS21}.
Capturing different degrees of complexity with a single simulator is a significant challenge~\cite{TestChallenges21},
we envision that using multiple simulators for different phases of the development
could be a viable solution.
Orthogonally to scale,
the dynamicity of \acp{ccps},
in which violations of the nominal states are generally unpredictable,
is a major challenge for simulation~\cite{TestChallenges21}:
generating
critical scenarios automatically is another interesting challenge ahead.

\subsubsection*{Test Cases: Inputs.}
\acp{ccps} are made of (possibly very large) groups of computational entities,
which require multiple levels of testing, as multiagent systems \cite{TestMAS21}.
We can think of (at least) the following levels: single entity; neighbouring entities; global system.
The main challenges are posed by the third level,
since the number of possible system states is huge,
and in the test cases it is necessary to include states that trigger all the relevant global system behaviour,
e.g., an important self-adaptation capability \cite{TestingCAS15}.
The decision about which global system states should be actually brought out by test cases is a difficult trade-off between keeping the number of such states reasonably small,
and covering all the relevant situations \cite{TestASSurv21}.
In order to find a good trade-off in a (semi-)automatic way,
the availability of a formal system model and/or of formal requirements is a promising direction to be pursued.
\subsubsection*{Test Cases: Outcomes.}
The nature of \acp{ccps} poses challenges also for the determination of the test outcome,
in terms of the equality between an \emph{expected} result (test oracle),
and the \emph{actual} result of applying certain inputs to a certain system state.
The (automatic) computation of the expected result requires the availability of a formal specification,
similarly as the determination of relevant test inputs discussed above \cite{TestASSurv21}.
However, also the determination of the actual result is not trivial, for at least two reasons.
First, the global outcome of the test must be determined
as the value of a formula that must be computed
by potentially taking into account the local states of a large number of system entities;
in order to compute it in a distributed way,
techniques such as distributed \ac{rv} may be adopted~\cite{CTLRV22}.
Second, the resulting global state may take some time to materialize
due to delays introduced by the communications between the \ac{ccps} entities.
Even when a \ac{ccps} behaviour is self-stabilising \cite{DBLP:journals/tomacs/ViroliABDP18}, then,
it is key to allow enough time for the stabilisation before finalising the test result.

\subsubsection*{Monitoring and runtime verification.}
As the runtime execution of a \ac{ccps}s is subject to external and often unpredictable influences, a-priori \ac{se} methods are usually not sufficient to guarantee the desired outcomes. Therefore, they have to be complemented by runtime monitoring approaches, both as separate tools flagging when manual intervention is needed, and as integrated parts of the system triggering automatic reactions. However, several challenges arise when performing runtime monitoring on \ac{ccps}. A first challenge is designing appropriate specification logics, that need to balance two opposite forces: expressiveness and simplicity. On expressiveness, a logic should capture enough relevant properties of systems situated in space and evolving over time. On simplicity, it should allow an intuitive understanding to enhance its usability, also among domain experts not proficient in modal logics. This tension has led to many different logics being developed so far \cite{DBLP:journals/iotj/MaBLSF21,DBLP:journals/lmcs/NenziBBL22,CTLRV22,DBLP:journals/jss/AudritoCDSV21}, and leaves room for improvement in future works.

A further challenge is computing monitor results from \ac{ccps} data. Some approaches such as  have been restricted so far to \emph{offline monitoring}, performed remotely on a complete trace of events gathered from the whole historical data of a \ac{ccps}, limiting their applicability in real-time \cite{DBLP:journals/lmcs/NenziBBL22}. Other approaches allow for \emph{centralised online monitoring} \cite{DBLP:journals/iotj/MaBLSF21}, still requiring to gather all events on a single remote computer, but allowing to compute monitor results in real-time from such event traces. Limited research has been carried out so far on \emph{decentralised monitoring} \cite{CTLRV22,DBLP:journals/jss/AudritoCDSV21}, in which monitors are integrated within the monitored application and run on the same distributed devices. Each device computes its own view of the monitored properties according to the knowledge available to it, without need for a central coordinator. Although this approach allows for greater resilience and faster reaction times of monitors, it poses additional challenges that have so far limited the monitored logic expressiveness. Future works could improve the current expressiveness limits in this context.

\subsection{Integration with Humans}

An important characteristic of \acp{ccps} is that they typically involve both physical components (such as sensors, robots, drones, ...) as well as humans.
The interaction between humans and the other cyber-physical entities can be either direct (e.g., through voice and gestures directly captured with microphones and cameras), or indirect, with the human interacting through wearable devices.
Two scenarios (of increasing complexity) of integration between humans and \acp{ccps} can be envisioned and pose peculiar challenges.

\subsubsection*{Humans controlling \acp{ccps}.} Humans and physical devices interact with strict, rigid asymmetric roles. We can consider humans as external, although strictly connected with the \ac{ccps}.
In this scenario, the humans typically request services from the \acp{ccps}, and/or perform some control over them \cite{HRIAffect20}. In a variation of the scenario, instead of commands, the humans may provide relevant inputs by carrying wearables or smartphones, e.g. in healthcare \cite{MASRehab19}.

One crucial aspect in this kind of systems is the cognitive load affecting the human(s) in charge of controlling the \ac{ccps}, especially when the latter contains many elements, as in a swarm \cite{HRISwarm22}.
Another important aspect is the communication modality between humans and devices (e.g., robots), which can involve speech, gesture, haptics (if humans and devices are physically close), or screen-based UIs and Virtual Reality (if the humans and devices are remote) \cite{HRISurvey23}.

The main challenges %ahead
 for this kind of scenario are thus mainly related with \ac{hmi}, and in particular multi-modal interfaces that improve the human user effectiveness and comfort.
A fundamental issue that is related with \ac{hmi} is the level of autonomy of the \ac{ccps} w.r.t. the human: the \emph{right} level of autonomy should be ideally adjusted dynamically to maximize the performance of the human high-level control over the \ac{ccps} \cite{ADJAUTSurvey19}.

\subsubsection*{Humans as \ac{ccps} Entities.} Humans and physical devices act as a single \ac{ccps}.
In this kind of scenario, there isn't a rigid hierarchy where humans give (high-level) commands and \ac{ccps} entities collectively execute them; instead, the roles of humans and robots are equal, or at least overlapping, and (partially) interchangeable.
Interesting applications of these hybrid \acp{ccps} include industrial processes \cite{MASIndustry17,MASFactory21,HRIAircraft24}, as well as rescue missions \cite{HRIMission18}.
In order to fully take part into the \ac{ccps}, humans typically have to carry some devices (smartphones, wearables, ...) used to provide inputs and outputs from/to the human.
The main challenge posed by this kind of systems is that they should run collective algorithms that take into account and exploit the differences between humans and artificial entities (e.g., intelligence, or speed), while at the same time considering them as (partially) interchangeable elements of the system.
This means that the system must have a more or less sophisticated model of its entities, and adapt its decisions to the involved entities.

\subsection{Learning collective behaviour}\label{ssec:chall:learning}

Despite years of research into synthesizing collective behaviours,
 finding an effective and scalable method to address the challenges associated with learning in such systems remains a complex issue~\cite{DBLP:journals/aamas/Hernandez-LealK19}.
The intricacy of multi-agent systems escalates significantly as the number of agents increases~\cite{DBLP:journals/corr/abs-1810-05587,DBLP:journals/air/DuD21};
 this amplifies the complexity of the system's dynamics,
 complicating the process of discerning the influence of individual actions on the collective outcome.
Below, we delineate some paramount challenges associated with learning in collective behaviour.

\subsubsection*{Multi-agent Credit Assignment}
A paramount issue within multi-agent systems is the \emph{multi-agent credit assignment} challenge,
 as highlighted by Nguyen et al.~\cite{DBLP:conf/nips/NguyenKL18}.
This problem revolves around the precise identification and rewarding of individual agent actions that contribute positively towards the collective goal.
Techniques such as COMA and difference rewards~\cite{DBLP:conf/atal/TumerA07} have been developed to tackle this challenge.
 However, these approaches often entail a significant computational burden,
 rendering them less viable for deployment in large-scale systems.
The core strategy of these methods involves calculating rewards for agents by assessing the impact of their actions on the collective outcome---
 a process that requires simulating the system's state with and without the agent's action to find the differential effect~\cite{DBLP:conf/atal/AgoginoT04}.
 This dual-simulation approach, while conceptually appealing for its accuracy in attributing credit,
 is notably resource-intensive and may not scale efficiently with the complexity or size of the system~\cite{DBLP:conf/aaai/FoersterFANW18}.

\subsubsection*{Online Learning} In such systems,
 agents should learn in real-time,
 adapting to changes in the environment and the behaviour of other agents
 without leveraging a central trainer.
However, unlike single-agent systems, where the environment typically exhibits stationary characteristics,
 online learning in multi-agent systems introduces a non-stationary environment~\cite{DBLP:journals/corr/abs-1810-05587,DBLP:journals/air/DuD21,DBLP:journals/tsmc/BusoniuBS08}.
 This complexity makes it difficult to determine whether changes are due to environmental responses or agent behaviour change,
 potentially leading to undesirable emergent behaviours and loss of control over collective learning~\cite{DBLP:journals/corr/abs-2011-00583,DBLP:journals/apin/OroojlooyH23}.
 Thus, moving from the current state of the art, which mainly focuses on offline learning with a centralised training/decentralised execution approach~\cite{DBLP:journals/air/GronauerD22,DBLP:journals/ijon/KraemerB16}, to decentralised online learning is a crucial challenge to be further investigated~\cite{DBLP:journals/air/DuD21}.

\subsubsection*{Transfer learning}
Another aspect of learning involves the \emph{transfer} of knowledge from one domain to another.
 The capability to apply knowledge acquired in one context to another, but similar, context presents a considerable challenge~\cite{DBLP:journals/jair/SilvaC19}.
Even if several solutions have been proposed so far to address this issue~\cite{DBLP:conf/ewrl/BoutsioukisPV11},
 the complexity of transfer learning in \acp{ccps} underscores the need for novel approaches to facilitate knowledge transfer across contexts effectively.
Indeed, in multiagent systems there are several axes is to be handled, such as the communication between agents, the different roles of agents,
the different capabilities of agents,
and the different environments in which agents operate~\cite{DBLP:journals/jair/SilvaC19}.
In single agent learning, a novel trend consist in leveraging large pre-trained models (i.e., \emph{foundational} models) to facilitate learning in new tasks~\cite{DBLP:journals/tmlr/ReedZPCNBGSKSEBREHCHVBF22}.
The adoption of foundational models in \acp{ccps} could mark a shift toward more efficient and effective learning mechanisms,
  suggesting a promising avenue for future exploration.

\section{Conclusion}\label{sec:conc}

In this paper, we presented challenges, current research efforts, and future research directions for a collective computing paradigm in software engineering.
We showed the relevance of this paradigm given the growing interest in large-scale cyber-physical ecosystems. % (cf. swarms, computing infrastructures, smart cities).
By reviewing key themes and relevant contributions in \acp{ccps} engineering, we categorised current research efforts on both techniques, abstractions, and tools.
From this, we identified the challenges ahead that will require innovative solutions.
These include the handling of homogeneity and heterogeneity; the development of DSLs, middleware and tools; the bridging of global and local perspectives in system analysis, design, and implementation; the integration with humans; and collective learning approaches.
Addressing these challenges will be essential for unlocking the full potential of \acp{ccps} and advancing the state of the art in software engineering cyber-physical ecosystems research.

\section*{Acknowledgements}

This work has been partially supported by the Italian MUR PRIN 2020 Project ``COMMON-WEARS'' (2020HCWWLP).

\bibliographystyle{ACM-Reference-Format}
\bibliography{bibliography}

\end{document}